%% file: HIG-11-024_temp.tex
\pdfoutput=1

\documentclass[11pt,twoside,a4paper,cmspaper,final,collab]{cms-tdr}
\def\svnVersion{101355:102181}\def\cmsCernNo{2012-018}\def\cmsCernDate{\today}\def\cmsMessage{Submitted to Physics Letters B}

\begin{document}\cmsNoteHeader{HIG-11-024}

\hyphenation{had-ron-i-za-tion}
\hyphenation{cal-or-i-me-ter}
\hyphenation{de-vices}

\RCS$Revision: 101373 $
\RCS$HeadURL: svn+ssh://alverson@svn.cern.ch/reps/tdr2/papers/HIG-11-024/trunk/HIG-11-024.tex $
\RCS$Id: HIG-11-024.tex 101373 2012-02-05 17:04:40Z alverson $
\newlength\cmsFigWidth\setlength\cmsFigWidth{0.49\textwidth}
\ifthenelse{\boolean{cms@external}}{\providecommand{\cmsLeft}{top}}{\providecommand{\cmsLeft}{left}}
\ifthenelse{\boolean{cms@external}}{\providecommand{\cmsRight}{bottom}}{\providecommand{\cmsRight}{right}}

\newcommand{\Hi}{\PH}
\newcommand{\W}{\PW}
\newcommand{\Wjets}{\ensuremath{\PW+\mathrm{jets}}}
\newcommand{\Zjets}{\ensuremath{\cPZ+\mathrm{jets}}}
\newcommand{\WW}{\PWp{}\PWm}
\newcommand{\ZZ}{\ensuremath{\cPZ\cPZ}}
\newcommand{\WZ}{\ensuremath{\PW\cPZ}}
\newcommand{\El}{\Pe}
\newcommand{\Elp}{\Pep}
\newcommand{\Elm}{\Pem}
\newcommand{\Elpm}{\ensuremath{\Pe^{\pm}}}
\newcommand{\Elmp}{\ensuremath{\Pe^{\mp}}}
\newcommand{\M}{\Pgm}
\newcommand{\Mp}{\Pgmp}
\newcommand{\Mm}{\Pgmm}
\newcommand{\Mmp}{\ensuremath{\Pgm^{\mp}}}
\newcommand{\Tau}{\Pgt}
\newcommand{\Nu}{\cPgn}
\newcommand{\Nubar}{\cPagn}
\newcommand{\Lep}{\ensuremath{\mathrm{\ell}}}
\newcommand{\Lepp}{\ensuremath{\mathrm{\ell}^{+}}}
\newcommand{\Lepm}{\ensuremath{\mathrm{\ell}^{-}}}
\newcommand{\Lprime}{\ensuremath{\Lep^{\prime}}}
\newcommand{\PP}{\Pp{}\Pp}
\newcommand{\PPbar}{\Pp{}\Pap}
\newcommand{\qq}{\Pq{}\Pq}
\newcommand{\To}{\ensuremath{\rightarrow}}

\newcommand{\mHi}{\ensuremath{m_{\PH}}}
\newcommand{\mW}{\ensuremath{m_{\PW}}}
\newcommand{\mZ}{\ensuremath{m_{\cPZ}}}
\newcommand{\mll}{\ensuremath{m_{\Lep\Lep}}}

\newcommand{\ptveto}{\ensuremath{\pt^\text{veto}}}
\newcommand{\ptl}{\ensuremath{p_\perp^{\Lep}}}
\newcommand{\ptlmax}{\ensuremath{\pt^{\Lep,\mathrm{max}}}}
\newcommand{\ptlmin}{\ensuremath{\pt^{\Lep,\mathrm{min}}}}
\newcommand{\met}{\MET}
\newcommand{\delphill}{\ensuremath{\Delta\phi_{\Lep\Lep}}}
\newcommand{\deletall}{\ensuremath{\Delta\eta_{\Lep\Lep}}}
\newcommand{\delphimetl}{\ensuremath{\Delta\phi_{\met\Lep}}}
\newcommand{\delphimetll}{\ensuremath{\Delta\phi_{\met\Lep\Lep}}}
\newcommand{\Et}{\ET}
\newcommand{\delR}{\ensuremath{\Delta R}}
\newcommand{\Eta}{\ensuremath{\eta}}
\newcommand{\GAMMA}{\Pgg}

\newcommand{\effsig}{\ensuremath{\varepsilon_{\text{bkg}}^{\mathrm{S}}}}
\newcommand{\effnorm}{\ensuremath{\varepsilon_{\text{bkg}}^{\mathrm{N}}}}
\newcommand{\Nsig}{\ensuremath{N_{\text{bkg}}^{\mathrm{S}}}}
\newcommand{\Nnorm}{\ensuremath{N_{\text{bkg}}^{\mathrm{N}}}}

\newcommand{\dyee}{\ensuremath{\Z/\GAMMA^*\to \Pep\Pem}}
\newcommand{\dymm}{\ensuremath{\Z/\GAMMA^*\to\Pgmp\Pgmm}}
\newcommand{\dytt}{\ensuremath{\Z/\GAMMA^* \to\Pgt^+\Pgt^-}}
\newcommand{\dyll}{\ensuremath{\Z/\GAMMA^*\to \ell^+\ell^-}}
\newcommand{\zee}{\ensuremath{\Z\to \Pep\Pem}}
\newcommand{\zmm}{\ensuremath{\Z\to\Pgmp\Pgmm}}
\newcommand{\ztt}{\ensuremath{\Z\to\Pgt^+\Pgt^-}}
\newcommand{\zll}{\ensuremath{\Z\to \ell^+\ell^-}}
\newcommand{\ppww}{\ensuremath{\Pp\Pp \to \PWp\PWm}}
\newcommand{\wwlnln}{\ensuremath{\PWp\PWm\to \ell^+\cPgn \ell^-\cPagn}}
\newcommand{\ww}{\PW\PW}
\newcommand{\wwpm}{\PWp\Pwm}
\newcommand{\hww}{\Hi\to\WW}
\newcommand{\wz}{\PW\cPZ}
\newcommand{\zz}{\cPZ\cPZ}
\newcommand{\wgamma}{\ensuremath{\W\GAMMA}}
\newcommand{\wjets}{\ensuremath{\PW+\text{jets}}}
\newcommand{\tw}{\cPqt\PW}

\newcommand{\ee}{\Pe\Pe}
\newcommand{\emu}{\Pe\Pgm}
\renewcommand{\mm}{\Pgm\Pgm}

\newcommand{\usedLumi}{4.6\fbinv}
\newcommand{\usedLumiWithSyst}{\ensuremath{4.6\pm0.2\fbinv}}

\cmsNoteHeader{HIG-11-024} 
\title{Search for the standard model Higgs boson decaying to \PWp{}\PWm\ in the
fully leptonic final state in pp collisions at $\sqrt{s}$ = 7 $\TeV$}

\date{\today}

\abstract{A search for the standard model Higgs boson decaying to
\PWp{}\PWm\ in pp collisions
at $\sqrt{s} = 7\TeV$ is reported. The data are collected at the LHC
with the CMS detector, and correspond to an integrated luminosity of 4.6\fbinv.
The \PWp{}\PWm\ candidates
are selected in events with two charged leptons and large missing
transverse energy. No
significant excess of events above the standard model background expectations is
observed, and upper limits on the Higgs boson production relative
to the standard model Higgs expectation are derived.
The standard model Higgs boson is excluded in the mass range 129--270\GeV
at 95\% confidence level.}

\hypersetup{%
pdfauthor={CMS Collaboration},%
pdftitle={Search for the standard model Higgs boson decaying to a W pair in the
fully leptonic final state in pp collisions at sqrt(s) = 7 TeV},%
pdfsubject={CMS},%
pdfkeywords={CMS, physics, Higgs}}

\maketitle 

\section{Introduction}
One of the open questions in the standard model (SM) of particle
physics~\cite{SM1,SM2,SM3} is the origin of the masses of fundamental
particles. Within the SM, vector boson masses arise by the
spontaneous breaking of electroweak symmetry by the
Higgs field~\cite{Higgs1, Higgs2, Higgs3,Higgs4, Higgs5, Higgs6}.
The existence of the associated field quantum, the Higgs boson, has yet to be
established experimentally. The discovery or the exclusion of the SM
Higgs boson is one of the central goals of the CERN Large Hadron Collider (LHC)
physics program.

Direct searches at the CERN \Pep{}\Pem\ LEP collider set a limit on
the Higgs boson mass $m_{\PH} > 114.4\GeV$ at 95\% confidence level
(CL)~\cite{LEPHIGGS}. Precision electroweak data
constrain the mass of the SM Higgs boson to
be less than 158\GeV at 95\% CL~\cite{EWK}.
The SM Higgs boson is excluded at 95\% CL by the
Tevatron collider experiments in the mass range 162--166\GeV~\cite{TEVHIGGS_2010},
and by the ATLAS experiment in the mass ranges 145--206, 214--224,
340--450\GeV~\cite{AtlasHWW,AtlasHZZ4l,AtlasHZZ2l2n}.
The $\Hi \to \WW \to 2\ell 2\cPgn$ final state, where $\ell$ is a charged lepton
and $\nu$ a neutrino, was first proposed as a clean channel at the LHC in~\cite{dittmar}.
A previous search for the Higgs boson at the LHC in this final state
was published  by the Compact Muon Solenoid (CMS) collaboration
with 36\pbinv of integrated luminosity~\cite{HWW2010}.
This search is performed over the mass range 110--600\GeV, and the data
sample corresponds to \usedLumiWithSyst\ of integrated luminosity
collected in 2011 at a center-of-mass energy of 7\TeV.
A similar search was conducted by the ATLAS collaboration~\cite{AtlasHWW}.

\section{CMS detector and simulation}
\label{sec:cms}
In lieu of a detailed description of the CMS detector~\cite{CMSdetector},
which is beyond the scope of the letter, a synopsis of the main components follows.
The superconducting solenoid occupies the
central region of the CMS detector, providing an axial magnetic
field of 3.8\unit{T} parallel to the beam direction. Charged particle trajectories
are measured by the silicon pixel and strip tracker, which cover a
pseudorapidity region of $|\eta| < 2.5$. Here, the pseudorapidity is defined as
$\eta=-\ln{(\tan{\theta/2})}$, where $\theta$ is the polar angle of the trajectory of the particle
with respect to the direction of the counterclockwise beam. The crystal electromagnetic
calorimeter (ECAL) and the brass/scintillator hadron calorimeter (HCAL)
surround the tracking volume and cover \mbox{ $|\eta| < 3$}. The steel/quartz-fiber
Cherenkov calorimeter (HF) extends the coverage to $|\eta| < 5$. The muon system consists of
gas detectors embedded in the iron return yoke outside the solenoid, with a coverage of $|\eta| < 2.4$.
The first level of the CMS trigger system, composed of custom
hardware processors, is designed to select the most interesting events
in less than 3\mus, using information from the calorimeters and muon
detectors. The High Level Trigger processor farm further
reduces the event rate to a few hundred Hz before data storage.

The expected SM Higgs cross section is 10 orders of magnitude smaller than the LHC
inelastic cross section, which is dominated by QCD processes.
Selecting final states with two leptons and missing energy eliminates the bulk of the QCD events,
leaving non-resonant diboson production ($\PP \to \WW$, $\WZ$, $\wgamma$, $\ZZ$), Drell-Yan production (DY),
top production ($\ttbar$ and $\tw$), and $\Wjets$ and QCD multijet
processes, where at least one jet is misidentified as a lepton, as the
background sources.
Several Monte Carlo event generators are used to simulate the signal and background processes.
The \textsc{powheg 2.0} program~\cite{powheg} provides event samples for the $\hww$ signal
and the Drell-Yan, \ttbar, and $\tw$ processes.
The $\Pq\Paq\to\WW$ and $\Wjets$ processes are generated using the \textsc{madgraph 5.1.3}~\cite{madgraph}
event generator, the $\cPg\cPg \to \WW$ process using \textsc{gg2ww}~\cite{ggww}, and the
remaining processes using \textsc{pythia 6.424}~\cite{pythia}.
For leading-order generators, the default set of parton distribution functions
(PDF) used to produce these samples is \textsc{cteq6l}~\cite{cteq66}, while
\textsc{ct10}~\cite{ct10} is used for next-to-leading order (NLO) generators.
Cross section calculations~\cite{LHCHiggsCrossSectionWorkingGroup:2011ti}
at next-to-next-to-leading order (NNLO) are used for the $\hww$ process, while NLO
calculations are used for background cross sections.
For all processes, the detector response is simulated using a detailed
description of the CMS detector, based on the \textsc{geant4}
package~\cite{Agostinelli:2002hh}.
The simulated samples are reweighted to represent the distribution of number of
pp interactions per bunch crossing (pile-up) as measured in the data.

\section{\texorpdfstring{$\WW$}{WW} event selection}
\label{sec:ww_evtsel}

The search strategy for $\hww$ exploits diboson events where
both $\W$ bosons decay leptonically, resulting in
an experimental signature of two isolated,
high transverse momentum ($\pt$), oppositely charged leptons (electrons or muons) and large
missing transverse energy (mainly due to the undetected neutrinos), $\met$, defined
as the modulus of the negative vector sum of the transverse momenta
of all reconstructed particles (charged or neutral) in the event~\cite{PFT-09-001}.
To improve the signal sensitivity,
the events are separated into three mutually exclusive categories according to the jet
multiplicity: 2$\ell$ with $\met$ + 0~jets, 2$\ell$ with $\met$ + 1~jet, and
2$\ell$ with $\met$ + 2~jets. Events with more than 2 jets are not considered.

Furthermore, the search strategy splits signal candidates into
three final states denoted by: $\Elp\Elm$, $\Mp\Mm$, and
$\Elpm\Mmp$.
The bulk of the signal arises through direct \PW\ decays to charged
stable leptons of opposite charge, though the small contribution
proceeding through an intermediate $\Pgt$ lepton is implicitly included.
The events are selected by triggers which require
the presence of one or two high-$\pt$ electrons or muons.
The trigger efficiency for signal events is measured to be above 95\% in the $\Mp\Mm$
final state, and above 98\% in the $\Elp\Elm$ and $\Elpm\Mmp$ final states for a
Higgs boson mass ${\sim}130\GeV$. The trigger efficiencies increase with the Higgs boson mass.

Two oppositely charged lepton candidates are required, with $\pt >20\GeV$ for
the leading lepton ($\ptlmax$) and $\pt >10\GeV$ for the trailing lepton ($\ptlmin$).
To reduce the low-mass $\dyll$ contribution, the requirement on the trailing lepton $\pt$ is raised to 15 $\GeV$
for the $\Elp\Elm$ and $\Mp\Mm$ final states. This tighter requirement also
suppresses the $\Wjets$ background in these final states. Only electrons (muons) with $|\eta| <$2.5 (2.4) are considered in the analysis.
Muon candidates~\cite{muonpas} are identified using a selection similar
to that described in~\cite{HWW2010}, while
electron candidates are selected using a multivariate
approach, which exploits correlations between the selection variables
described in~\cite{egmpas} to improve identification performance.
The lepton candidates are required to originate from the primary vertex of the
event, which is chosen as the vertex with highest $\sum \pt^2$, where the sum
is performed on the tracks associated to the vertex, including
the tracks associated to the leptons.
This criterion provides the correct assignment for the
primary vertex in more than 99\% of both signal and
background events for the pile-up distribution observed in the data.
Isolation is used to distinguish lepton candidates from \PW-boson decays from those
stemming from QCD background processes, which are usually immersed in hadronic activity.
For each lepton candidate, a $\Delta R \equiv\sqrt{(\Delta\eta)^2 + (\Delta\phi)^2}$ cone
of 0.3 (0.4) for muons (electrons) is constructed around the track direction at the event vertex.
The scalar sum of the transverse energy of each
particle reconstructed using a particle-flow algorithm~\cite{PFT-09-001} compatible with the
primary vertex and contained within the cone is calculated,
excluding the contribution from the lepton candidate itself. If this
sum exceeds approximately 10\% of the candidate $\pt$ the lepton is
rejected, the exact requirement depending on the lepton $\eta$, $\pt$ and
flavour.

Jets are reconstructed from calorimeter and tracker information using
the particle-flow technique~\cite{PFT-09-001,jetpas}, combining the information from
all CMS subdetectors to reconstruct each individual particle. The anti-$\mathrm{k_T}$
clustering algorithm~\cite{antikt} with distance parameter $\mathrm{
R}=0.5$ is used, as implemented in the \textsc{fastjet}
package~\cite{Cacciari:fastjet1,Cacciari:fastjet2}.
To correct for the contribution to the jet energy
due to the pile-up, a median energy density ($\rho$) is determined event by event.
Then the pile-up contribution to the jet energy is estimated as the product of $\rho$ and the area of the
jet and subsequently subtracted~\cite{Cacciari:subtraction} from the jet transverse energy $\Et$.
Jet energy corrections are also applied as a function of the jet $\Et$ and $\eta$~\cite{cmsJEC}.
Jets are required to have $\Et>30\GeV$ and $|\eta|<$5
to contribute to the event classification according to the jet multiplicity

In addition to high momentum isolated leptons and minimal jet activity, missing
energy is present in signal events but not in background.
In this analysis, a \textit{projected}~$\met$ variable, defined
as the component of $\met$ transverse to the nearest lepton if that lepton is within
$\pi/2$ in azimuthal angle, or the full $\met$ otherwise, is employed.
A cut on this observable efficiently rejects $\dytt$ background events, where the $\met$
is preferentially aligned with leptons, as well as $\dyll$ events with mismeasured
$\met$ associated with poorly reconstructed leptons or jets.
The $\met$ reconstruction makes use of event reconstruction via the particle-flow technique~\cite{PFT-09-001}.
Since the \textit{projected}~$\met$ resolution is degraded by pile-up,
a minimum of two different observables is used: the first includes all reconstructed particles in the
event~\cite{PFT-09-001}, while the second uses only the charged
particles associated with the primary vertex. For the same
cut value with the first observable, the $\dyll$ background doubles
when going from 5 to 15 pile-up events, while it remains approximately constant with
the second observable. The use of both observables exploits
the presence of a correlation between them in signal events with genuine
$\met$, and its absence otherwise, as in Drell-Yan events.

Drell-Yan background produces same-flavour lepton pairs ($\Elp\Elm$ and $\Mp\Mm$):
thus, the selection requirements designed to suppress this background are slightly
different for same-flavour and opposite-flavour ($\Elpm\Mmp$) events.
Same-flavour events must have \textit{projected}~$\met$ above about 40\GeV,
with the exact requirement depending on the number of reconstructed primary vertices ($N_\text{vtx}$)
according to the relation \textit{projected}~$\met >( 37 +N_\text{vtx}/ 2 )\GeV$.
For opposite-flavour events, the requirement is lowered to 20\GeV
with no dependence on the number of vertices.
These requirements remove more than 99\% of the Drell-Yan background.
In addition, requirements of a minimum dilepton transverse momentum ($\pt^{\ell\ell}$)
of 45\GeV
for both types and a minimum dilepton mass ($\mll$) of 20 (12)\GeV for same-
(opposite-) flavour events are applied.
Two additional selection criteria are applied only to the same-flavour events.
First, the dilepton mass must be outside a 30\GeV window centered on the $\Z$ mass,
and second, to suppress Drell-Yan events with the $\Z/\Pgg^*$ recoiling against a jet, the angle
in the transverse plane between the dilepton system and the leading jet must be less
than 165 degrees, when the leading jet has $\Et>15\GeV$.

To suppress the top-quark background, a \textit{top tagging} technique based
on soft-muon and b-jets tagging methods~\cite{btag1,btag2} is applied. The first
method is designed to veto events containing muons from b-quarks coming from the top-quark decay. The second
method uses b-jet tagging, which looks for tracks with large
impact parameter within jets. The algorithm is also applied in the
case of 0-jet bin, which can still contain jets with $\Et <30\GeV$.
The rejection factor for top-quark background is about two in the 0-jet
category and above 10 for events with at least one jet passing the selection criteria.

To reduce the background from $\WZ$ and $\ZZ$ production, any event
that has a third lepton passing the identification and isolation requirements is rejected.
This requirement rejects less than 0.1\% of the $\hww \to 2\ell2\nu$ events, while
it rejects 60\% of $\WZ$ and 10\% of the $\ZZ$ processes. After the $\met$ requirement
$\ZZ$ events are dominated by the $\ZZ \to 2\ell 2\cPgn$ process,
where there is no 3rd lepton. The $\wgamma$ production,
where the photon is misidentified as an electron, is reduced by more than 90\%
in the dielectron final state by $\GAMMA$ conversion rejection requirements.

\begin{figure}[htbp]
\begin{center}
  \includegraphics[width=\cmsFigWidth]{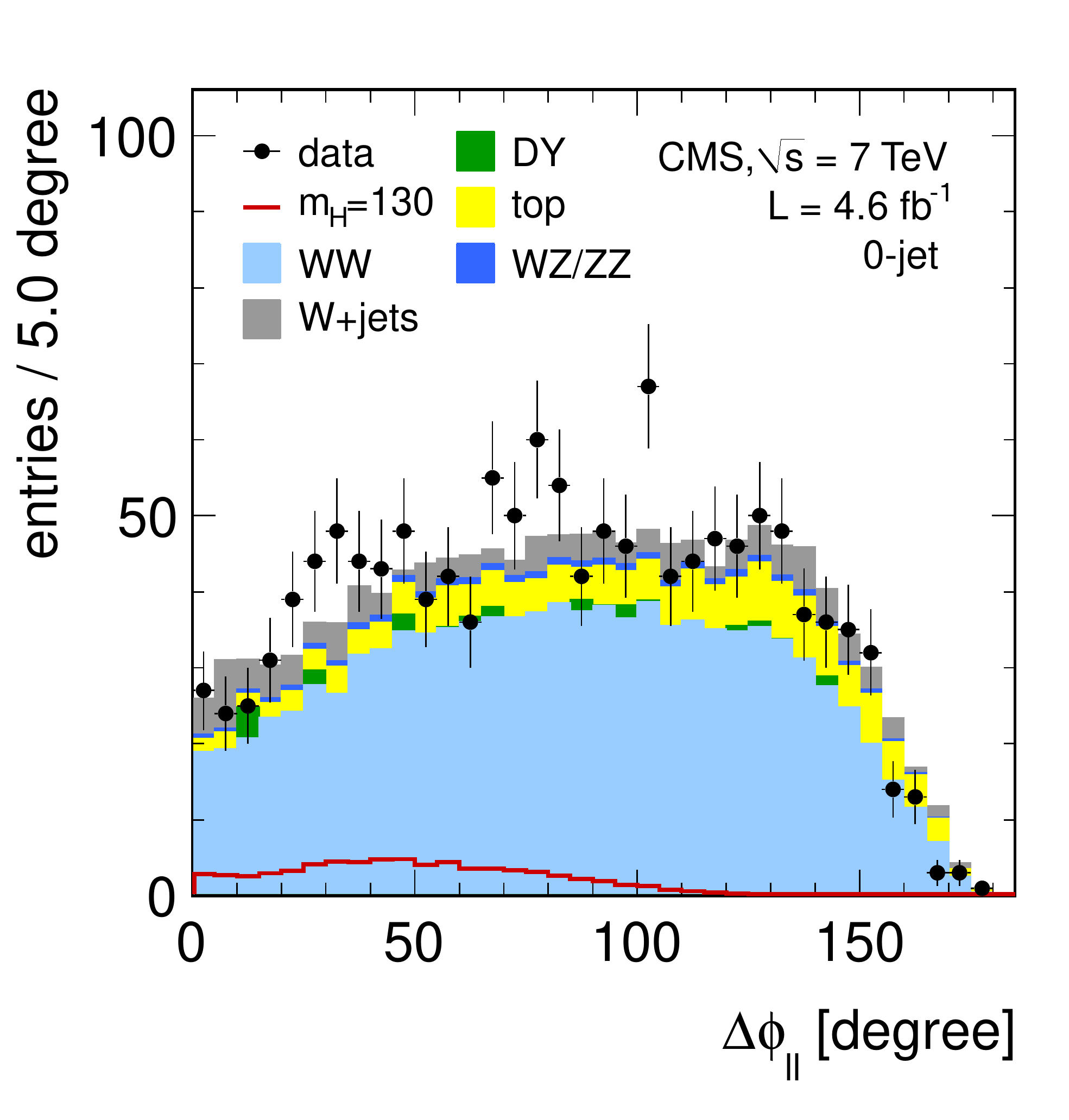}
  \includegraphics[width=\cmsFigWidth]{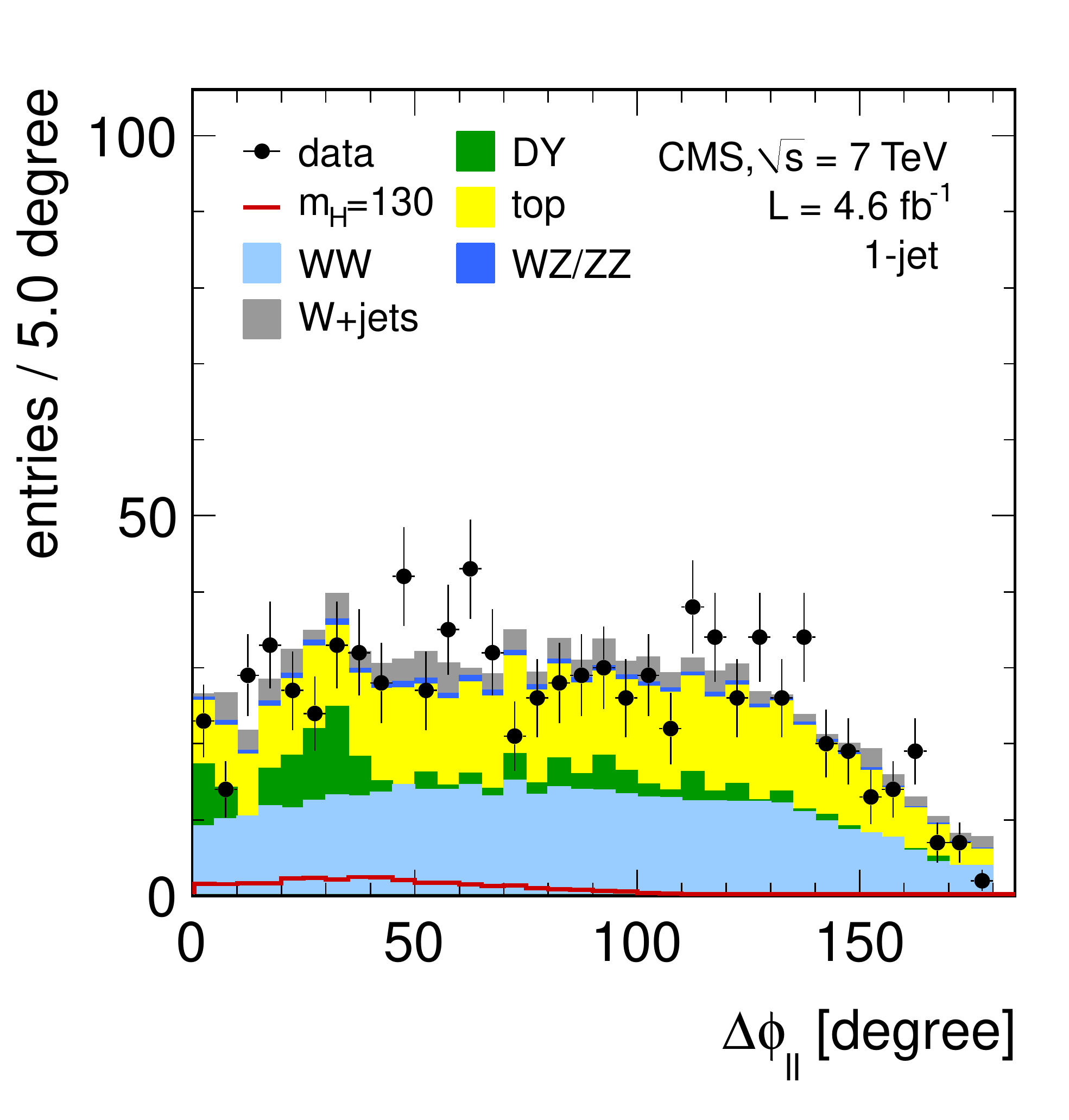}
	 \caption{Azimuthal angle difference between the two selected leptons
	 in the 0-jet (\cmsLeft) and 1-jet (\cmsRight) categories,
       for a $\mHi= 130\GeV$ SM Higgs boson and for the main
       backgrounds at the $\WW$ selection level.}  \label{fig:deltaphi}
\end{center}
\end{figure}

After applying all selection criteria described in this section, which is referred to
as the ``$\WW$ selection'', 1359, 909, and 703 events are obtained in data in the
0-jet, 1-jet, and 2-jet categories respectively. This sample is
dominated by non-resonant $\WW$ events. The efficiency at this stage for a
Higgs boson with $\mHi =130\GeV$ is about 5.5\%.
Figure~\ref{fig:deltaphi} shows the distributions of the azimuthal angle
difference ($\delphill$) between the two selected leptons
after the $\WW$ selection, for a SM Higgs boson with $\mHi=130\GeV$
and for backgrounds in the 0- and 1-jet categories. The scale of the figures
allows for comparing the background contributions between the 0-jet and the 1-jet channels.

\section{\texorpdfstring{$\Hi \to \WW$}{Higgs to WW} search strategy}
\label{sec:hww}

To enhance the sensitivity to a Higgs boson signal, two different analyses are
performed in the 0-jet and 1-jet categories, the first utilizing
a cut-based approach and the second using a multivariate
technique. Both cover a large range of Higgs boson masses.
As the kinematics of signal events change as a function of the Higgs mass,
separate optimizations are performed for different $m_{\PH}$ hypotheses.
Only the cut-based approach is applied to the 2-jet category,
as its relative impact on the sensitivity is limited with the current integrated luminosity.

\begin{figure}[htbp]
\begin{center}
  \includegraphics[width=\cmsFigWidth]{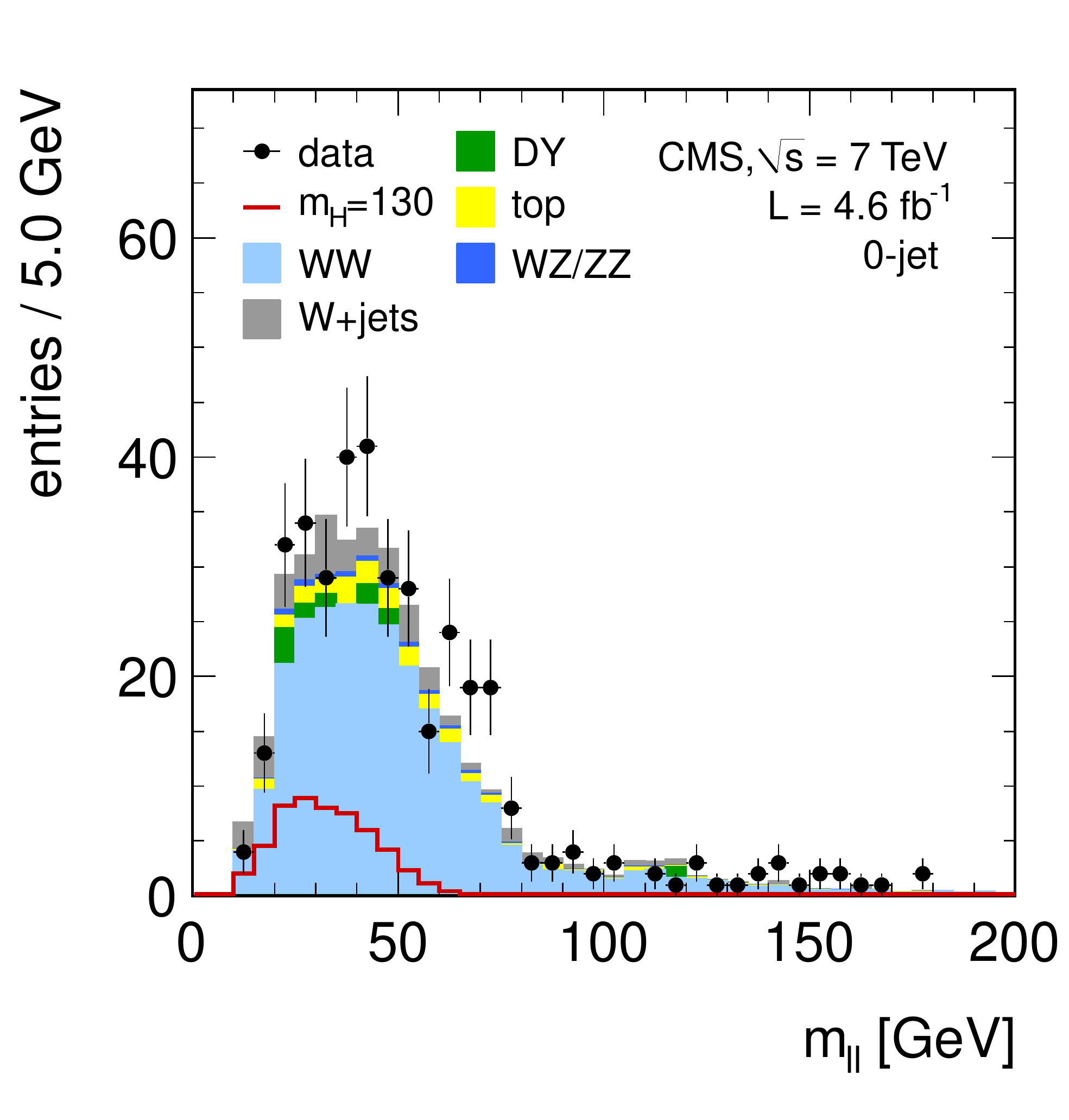}
  \includegraphics[width=\cmsFigWidth]{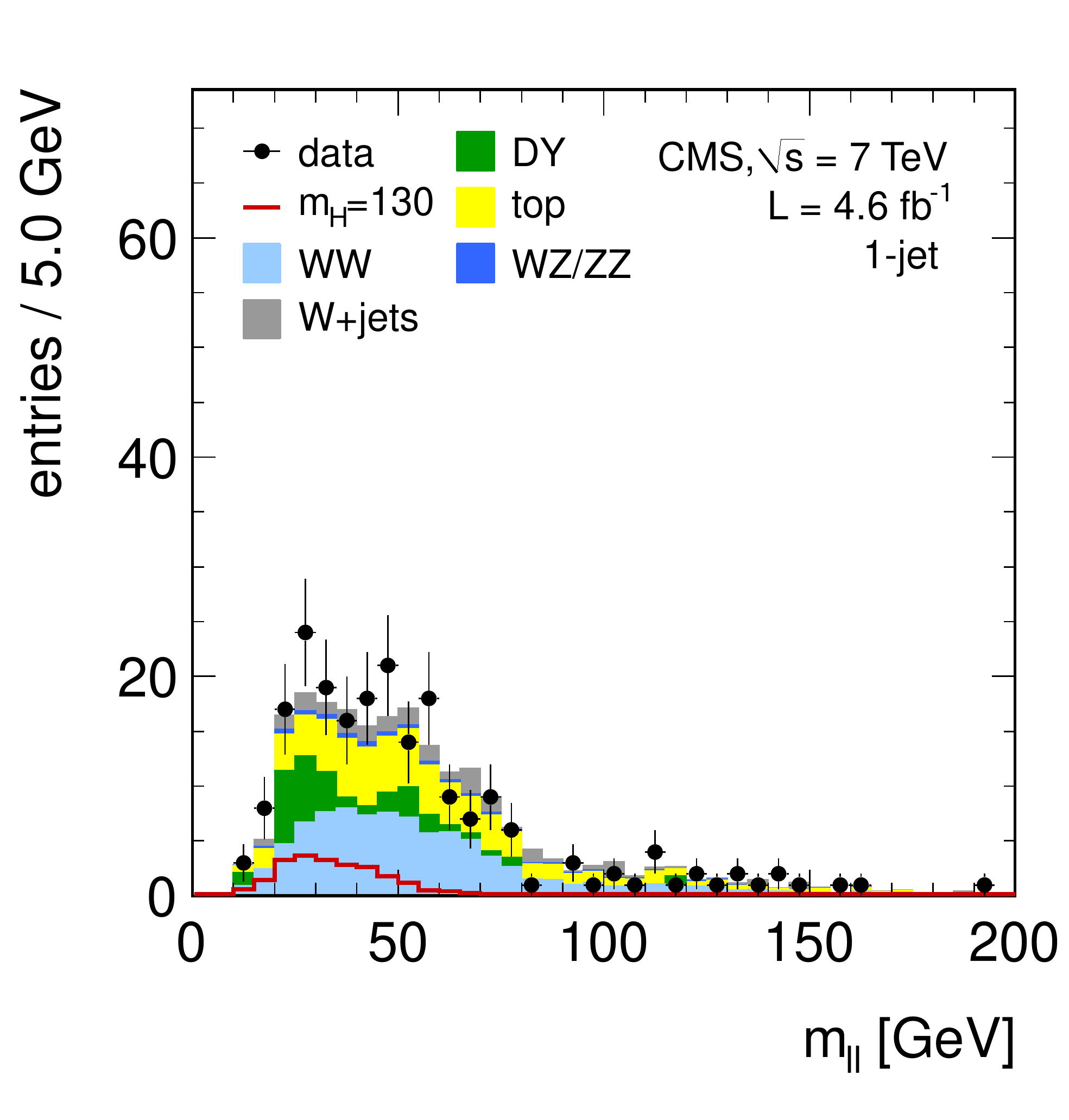}
	 \caption{Dilepton mass
	 in the 0-jet (\cmsLeft) and 1-jet (\cmsRight) categories,
       for a $\mHi=130\GeV$ SM Higgs boson and for the main backgrounds.
       The cut-based $\hww$ selection, except for the requirement on the dilepton mass itself, is applied.}  \label{fig:mllHWW}
\end{center}
\end{figure}

In the cut-based approach extra requirements, designed to optimize the
sensitivity for a SM Higgs boson, are placed
on $\ptlmax$, $\ptlmin$, $\mll$, $\delphill$ and
the transverse mass $m_\mathrm{T}$,
defined as $\sqrt{2 \pt^{\ell\ell} \met (1-\cos\delphimetll)}$, where $\delphimetll$
is the angle in the transverse plane between $\met$ and the transverse momentum of the
dilepton system.
The cut values, which are the same in both the 0- and 1-jet categories,
are summarized in Table~\ref{tab:cuts_analysis}. The $\mll$ distribution of
the two selected leptons in the 0-jet and 1-jet categories,
for a $\mHi=130\GeV$ SM Higgs hypothesis and for the main backgrounds,
are shown in Fig.~\ref{fig:mllHWW}.

\begin{table}[htbp]
  \begin{center}
  \caption{Final event selection requirements for the cut-based analysis in the 0-jet and 1-jet bins.
  The values of $\ptlmin$ in parentheses at low Higgs masses correspond to the requirements on the
  trailing lepton for the same-flavour final states.}
 {\small
      \setlength{\extrarowheight}{1pt}
  \begin{tabular} {|c|c|c|c|c|c|c|}
  \hline
$\mHi$       & $\ptlmax$ & $\ptlmin$ & $\mll$     & $\delphill$ & $m_\mathrm{T}$ \\  \hline
[$\GeVns$] & [$\GeVns$] & [$\GeVns$] & [$\GeVns$] & [\de]       & [$\GeVns$]             \\  \hline
           &   $>$     &   $>$     &   $<$      &  $<$        &    [,]                 \\  \hline

    120 & 20  &  10 (15) & 40  & 115 & [80,120]  \\
    130 & 25  &  10 (15) & 45  & 90  & [80,125]  \\
    160 & 30  &  25      & 50  & 60  & [90,160]  \\
    200 & 40  &  25      & 90  & 100 & [120,200] \\
    250 & 55  &  25      & 150 & 140 & [120,250] \\
    300 & 70  &  25      & 200 & 175 & [120,300] \\
    400 & 90  &  25      & 300 & 175 & [120,400] \\
  \hline
  \end{tabular}
  }
   \label{tab:cuts_analysis}
  \end{center}
\end{table}

In the multivariate approach a boosted decision tree (BDT)
is trained for each Higgs boson mass hypothesis~\cite{tmva} and jet category
to discriminate signal from background. In addition to the $\WW$ selection,
loose $\mHi$ dependent requirements on $\mll$ and
$m_\mathrm{T}$ are applied to enhance the signal-to-background ratio.

\begin{figure*}[htbp]
\begin{center}
   \includegraphics[width=\cmsFigWidth]{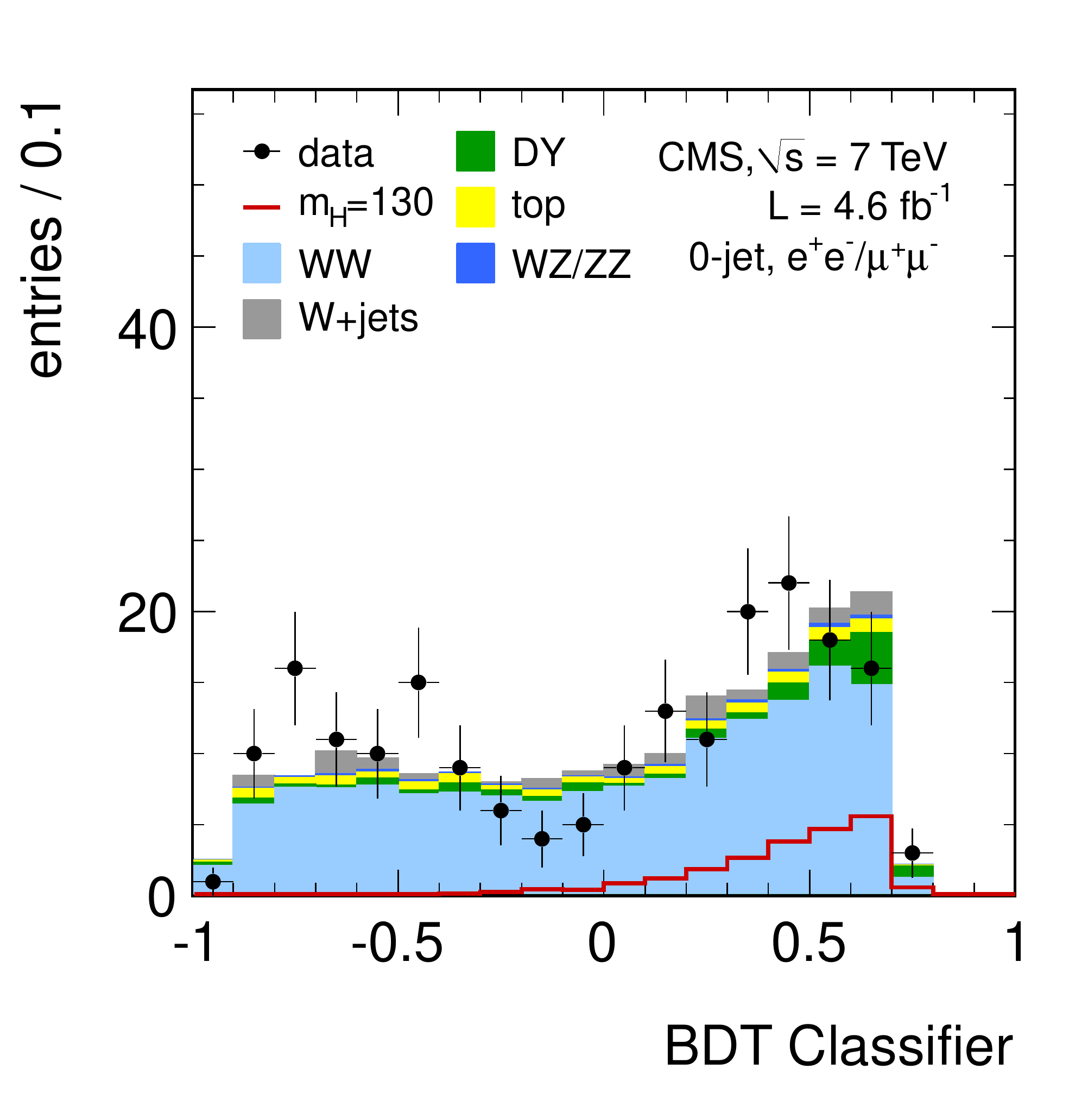}
   \includegraphics[width=\cmsFigWidth]{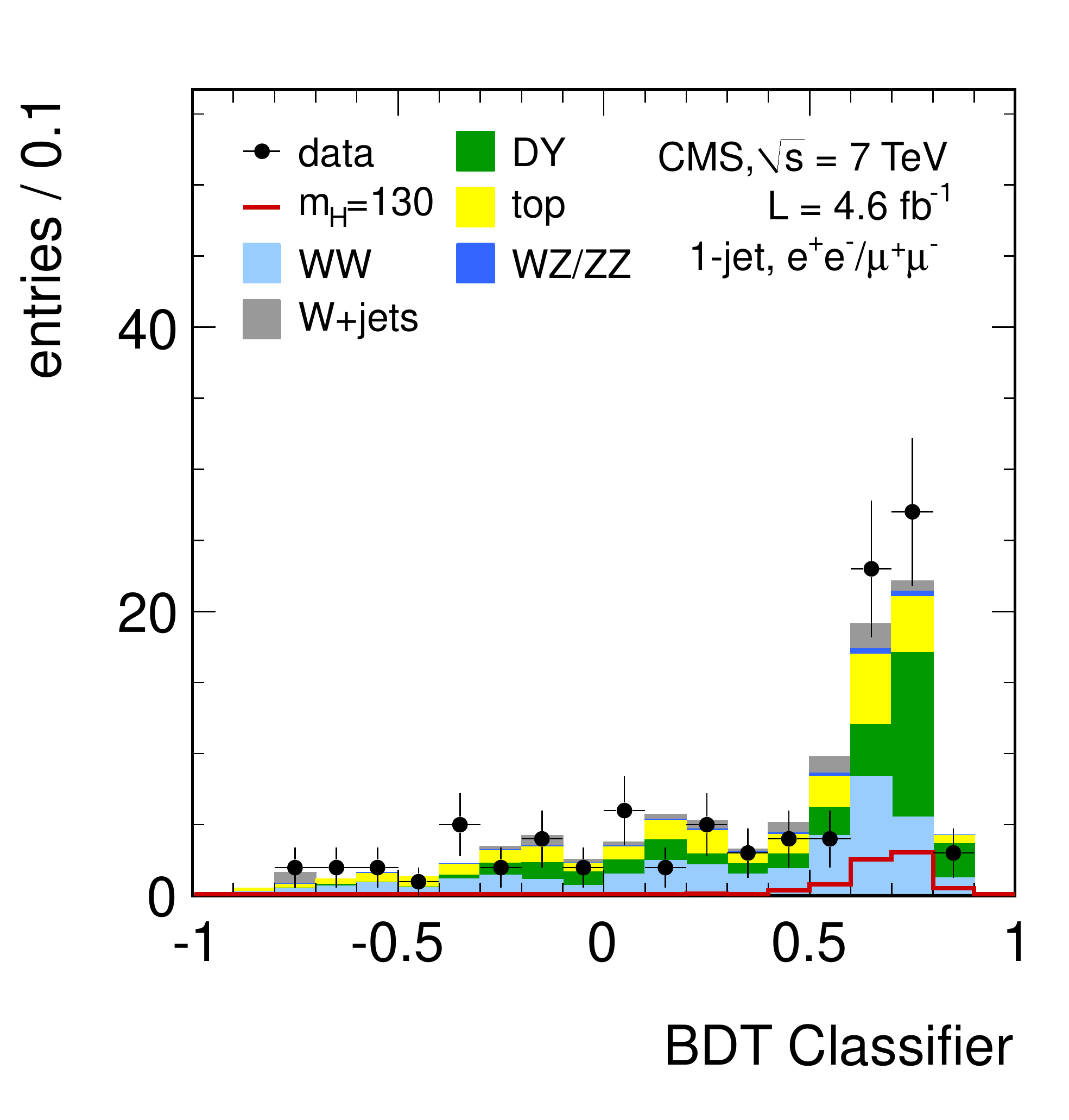}
   \includegraphics[width=\cmsFigWidth]{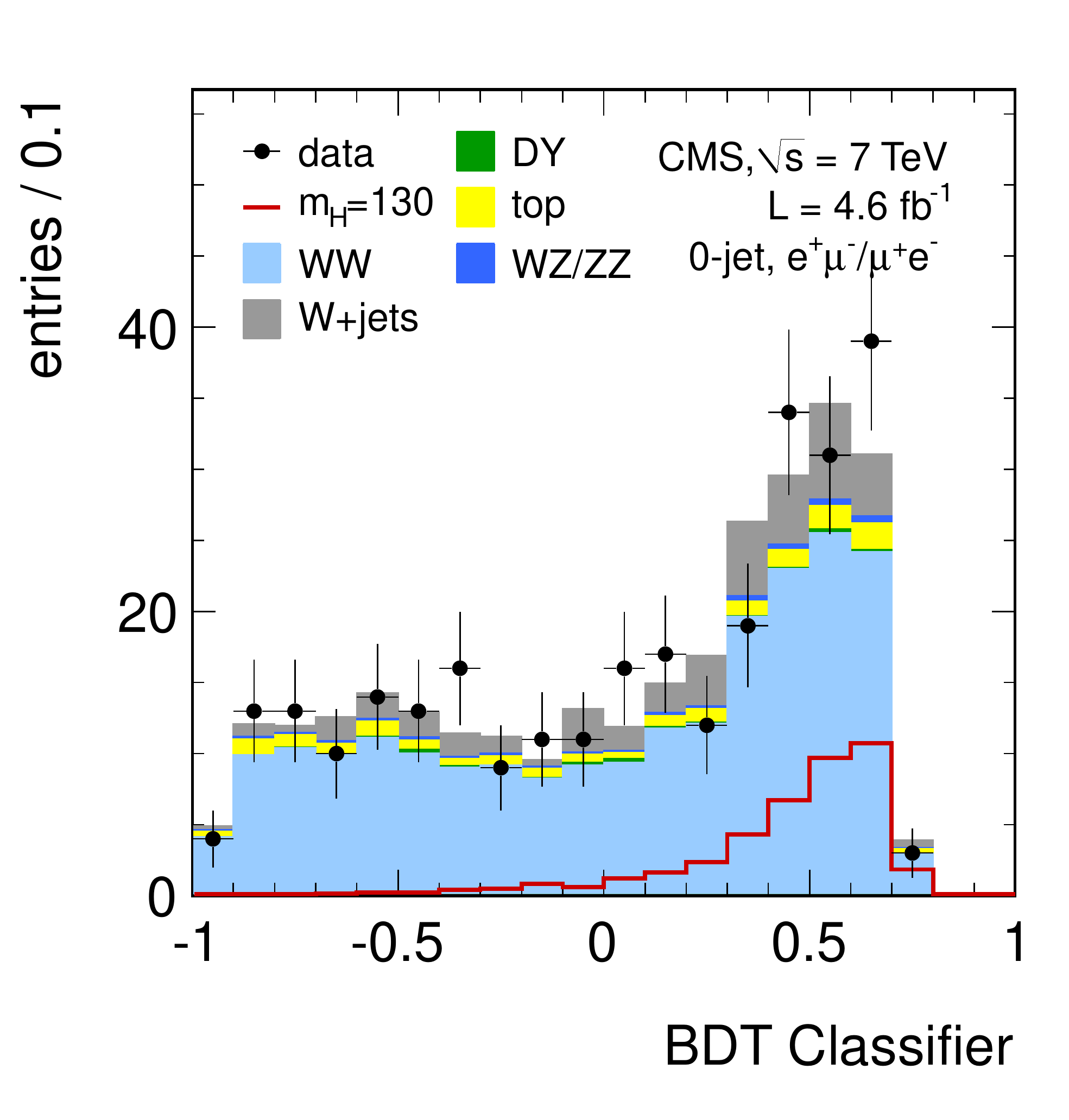}
   \includegraphics[width=\cmsFigWidth]{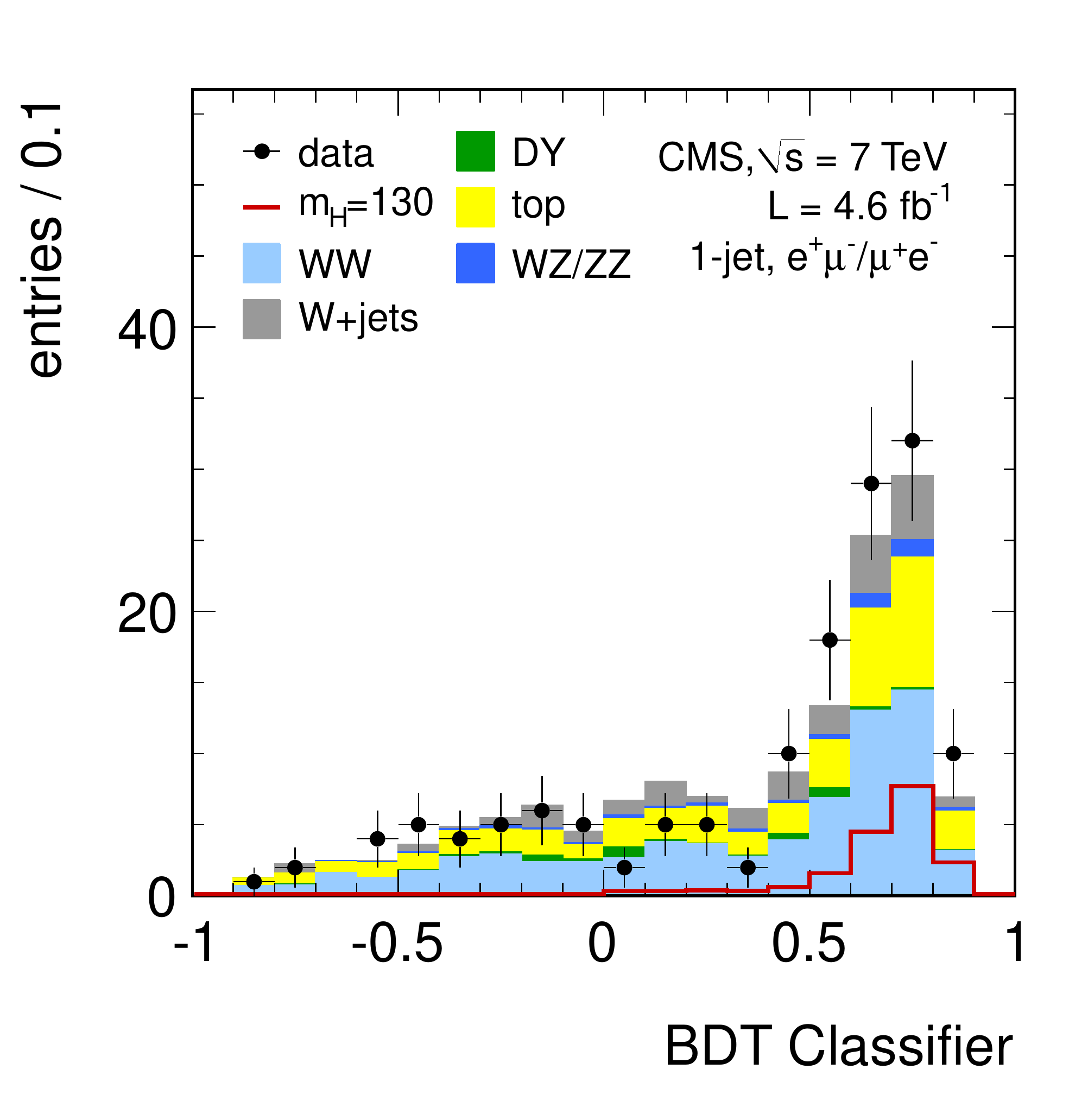}
       \caption{BDT classifier distributions for signal and background
events for a $\mHi=130\GeV$ SM Higgs boson and for the main backgrounds at the $\WW$ selection
level: (upper-left) 0-jet bin same-flavour final state, (upper-right) 1-jet bin same-flavour final state, (lower-left) 0-jet bin opposite-flavour
final state, (lower-right) 1-jet bin opposite-flavour final state.} \label{fig:histo_mva_130}
\end{center}
\end{figure*}

The multivariate technique
uses the following observables in addition to those used in the cut-based
analysis: $\Delta R_{\Lep\Lep}\equiv\sqrt{(\deletall)^2 + (\delphill)^2}$
between the leptons, the transverse mass of both lepton-$\met$ pairs, and finally
the lepton flavours.
The BDT training is performed using $\hww$ as signal and non-resonant
$\WW$ as background. Exhaustive studies demonstrate that
the inclusion of other processes does not improve
the performance, because the kinematic variables within the jet category and phase-space
region are quite similar among various background processes.
The BDT classifier distributions for $\mHi=130\GeV$ are shown in
Fig.~\ref{fig:histo_mva_130} for 0-jet and 1-jet categories. In the analysis, the binned
BDT distributions of Fig.~\ref{fig:histo_mva_130} are fitted to templates for the signal and
backgrounds BDT distributions.
The analysis is repeated using both a likelihood approach,
where the correlations among the variables are neglected, and a single
variable approach based on $\mll$.
We also perform an analysis using a Matrix Element
method as previously done in~\cite{Aaltonen:2008ec}, to compute the differential cross section
for signal and background hypotheses on an event-by-event basis.
At low masses of the Higgs boson, all approaches yield results
consistent with those from the BDT analysis,
which is chosen as default because of the superior sensitivity in the entire 110--600\GeV mass range.

The 2-jet category is mainly sensitive to the vector boson fusion (VBF) production mode
\cite{Ciccolini:2007jr, Ciccolini:2007ec, Arnold:2008rz}, whose
cross section is roughly ten times smaller than that for the
gluon-gluon fusion mode. The VBF channel with a different production mechanism offers the
possibility to test the compatibility of an eventual signal with the SM Higgs.
The VBF signal can be extracted
using simple selection criteria especially in the relatively low background environment
of the fully leptonic $\WW$ decay mode, providing additional search sensitivity.
The $\hww$ events from VBF production are characterized by a pair of
energetic forward-backward jets and very little hadronic activity in
the rest of the event.
Events passing the $\WW$ criteria are selected requiring $\pt>30\GeV$
for both leading jets, with no jets above this threshold
present in the pseudorapidity region between them.
To reject
the main background, which stems from top-quark decays, two additional requirements are applied
to the two jets, $j_1$ and $j_2$: $|\Delta\eta (j_1,j_2)| > 3.5$ and
$m_{j_1j_2} >450\GeV$. Finally, a $\mHi$ dependent requirement on
the high end of the dilepton mass is applied.

The selection with the requirements described in this section is referred to
as the ``Higgs selection'' for both the cut-based and the multivariate approaches.

\section{Background predictions}
\label{sec:backgrounds}

A combination of techniques are used to determine the contributions from the background
processes that remain after the Higgs selection.
Where feasible, background contributions are estimated directly from
the data itself, avoiding large uncertainties related to
the simulation of these sources. The remaining contributions
taken from simulation are small.

The $\Wjets$ and QCD multijet backgrounds arise from leptonic
decays of heavy quarks, ha\-drons misidentified as leptons, and electrons
from photon conversion. The estimate of these contributions is derived
directly from data using a control sample of events where one lepton
passes the standard criteria and the other does not, but satisfies a
relaxed set of requirements (``loose" selection), resulting in a
``tight-fail" sample.
The efficiency, $\epsilon_\text{loose}$, for a jet satisfying the loose
selection to pass the tight selection is determined using data from an
independent multijet event sample dominated by non-prompt leptons, and
parameterized as a function of $\pt$ and $\eta$ of such lepton. The background
contamination is then estimated using the events of the "tight-fail"
sample weighted by \mbox{$\epsilon_\text{loose}$/(1 - $\epsilon_\text{loose}$)}. The
systematic uncertainties stemming from the efficiency determination
dominate the overall uncertainty of this method, which is estimated
to be about 36\%.

The normalization of the top-quark background is estimated from data as well
by counting the number of top-tagged ($N_\text{tagged}$) events and applying the
corresponding top-tagging efficiency. The top-tagging efficiency
($\epsilon_\text{top tagged}$) is measured with a control sample dominated by
$\ttbar$ and $\tw$ events, which is selected by requiring a b-tagged jet.
The residual number of top events ($N_\text{not tagged}$) in the signal region is given
by: $N_\text{not tagged} = N_\text{tagged} \times
 (1-\epsilon_\text{top tagged})/\epsilon_\text{top~tagged}$.
The main uncertainty comes from the statistical
uncertainty in the control sample and from the systematic uncertainties related to
the measurement of $\epsilon_\text{top tagged}$. The uncertainty is about
25\% in the 0-jet category and about 10\% otherwise.

For the low-mass $\hww$ signal region, $m_{\PH} <200\GeV$,
the non-resonant $\WW$ contribution is estimated from data. This contribution
is measured using events with a dilepton mass larger than 100\GeV,
where the Higgs boson signal contamination is negligible, and a simulation
is used to extrapolate into the signal region. The total uncertainty is about 10\%.
For larger Higgs boson masses there is a large overlap between the
non-resonant $\WW$ and Higgs boson signal, and simulation is
used for the estimation.

The $\dyll$ contribution to the $\Elp\Elm$
and $\Mp\Mm$ final states is based on extrapolation from
the observed number of events with a dilepton mass
within $\pm7.5\GeV$ of the $\Z$ mass, where the residual
background on that region is subtracted, using $\Elpm\Mmp$ events.
The extrapolation to the signal region is performed using the simulation
and the results are cross-checked with data, using the same algorithm and
subtracting the background in the peaking region which is estimated from $\Elpm\Mmp$ events.
The largest uncertainty in the estimate is related to the statistical
uncertainty of the control sample and it is about 50\%.
The $\dytt$ contamination is estimated using $\dyee$ and $\Mp\Mm$
events selected in data, where the leptons are replaced with simulated
$\Tau$ decays, thus providing a better description of the experimental conditions
with respect to the full simulation of the process $\dytt$. The \textsc{tauola}~\cite{tauola} package is used in the
simulation of $\Tau$ decays to account for $\Tau$ polarization effects.

Finally, to estimate the $\wgamma^{*}$ background contribution coming
from asymmetric virtual photon decays~\cite{wgammastart}, where one lepton escapes
detection, the \textsc{madgraph} generator with dedicated cuts is
used. To obtain the normalization scale of the
simulated events a control sample of high purity $\wgamma^{*}$ events with three
reconstructed leptons is defined and compared to the simulation prediction. A
measured factor of about 1.6 with respect to the leading order
cross section is found.

Other minor backgrounds from $\WZ$, $\ZZ$ (when the two selected leptons come from
different bosons) and $\wgamma$ are estimated from simulation.
The $\wgamma$ background estimate is cross-checked in data using the events passing
all selection requirements, except that here the two leptons must have the
same charge; this sample is dominated by $\Wjets$ and $\wgamma$
events.

The number of estimated events for all processes after the $\WW$ selection
are summarized in Table \ref{tab:wwselection_all}. The number of events
observed in data for the cut-based selection, with the signal and
background predictions, are listed in Table~\ref{tab:hwwselection} for several
mass hypotheses.

\begin{table*}[htbp]
  \begin{center}
    \caption{Observed number of events and background estimates for
    an integrated luminosity of $\usedLumi$ after applying the $\WW$ selection requirements.
    Only statistical uncertainties on each estimate are reported.
    The $\dyll$ process corresponds to the dimuon and dielectron final states.}
     {
     \small
     \setlength{\extrarowheight}{1pt}
      \begin{tabular} {|c|c|c|c|c|c|c|}
\hline
          &   data & all bkg. & $\Pq\Paq \to \WW$ & $\Pg\Pg \to \WW$ &  $\ttbar+\cPqt\PW$   & $\Wjets$    \\
  \hline
  \hline
 0-jet & 1359 & 1364.8 $\pm$    9.3 &  980.6 $\pm$    5.2 &   58.8 $\pm$    0.7 &  147.3 $\pm$    2.5 &   99.3 $\pm$    5.0 \\
 1-jet &  909 &  951.4 $\pm$    9.8 &  416.8 $\pm$    3.6 &   23.8 $\pm$    0.5 &  334.8 $\pm$    3.0 &   74.3 $\pm$	4.6  \\
 2-jet &  703 &  714.8 $\pm$   13.5 &  154.7 $\pm$    2.2 &    5.1 $\pm$    0.2 &  413.5 $\pm$    2.7 &   37.9 $\pm$	3.6  \\
 \hline
 \hline
  \end{tabular}

  \begin{tabular} {|c|c|c|c|c|c|}
  \hline
       & $\WZ$/$\ZZ$ & $\dyll$ & $\wgamma^{(*)}$ & $\dytt$ \\
       \hline
       \hline
       0-jet & 33.0 $\pm$ 0.5  & 16.6 $\pm$    4.0 &   26.8 $\pm$    3.5 &    2.4 $\pm$    0.5 \\
       1-jet & 28.7 $\pm$ 0.5  & 39.4 $\pm$    6.4 &   13.0 $\pm$    2.6 &   20.6 $\pm$    0.4 \\
       2-jet & 15.1 $\pm$ 0.3  & 56.1 $\pm$   11.7 &   10.8 $\pm$    3.6 &   21.6 $\pm$    2.1 \\
       \hline
       \hline
       \end{tabular}
  }
   \label{tab:wwselection_all}
  \end{center}
\end{table*}

\begin{table*}[htbp]
  \begin{center}
  \caption{Observed number of events, background estimates and signal predictions
  for an integrated luminosity of $\usedLumi$ after applying the $\hww$ cut-based selection requirements.
  The combined statistical and experimental systematic uncertainties on the processes are reported.
  Theoretical systematic uncertainties are not quoted.
  The $\dyll$ process corresponds to the dimuon, dielectron and ditau final state.}
   \label{tab:hwwselection}
 {
 \ifthenelse{\boolean{cms@external}}{\small}{\scriptsize}
\setlength{\extrarowheight}{1pt}
\begin{tabular} {|c|c|c|c|c|c|c|c|c|}
  \hline
$\mHi$ & data & all bkg. & pp$\to \WW$ & top & $\Wjets$ & $\WZ+\ZZ+\wgamma^{(*)}$ & $\dyll$ & $\Hi \to \WW$ \\ \hline
\multicolumn{9}{|c|}{0-jet category} \\
  \hline
120 & $136$ & $136.7\pm12.7$ & $100.3\pm7.2$  & $6.7\pm1.0$  & $14.7\pm4.7$ & $6.1\pm1.5$ & $8.8\pm9.2$  & $15.7\pm0.8$  \\ \hline
130 & $193$ & $191.5\pm14.0$ & $142.2\pm10.0$ & $10.6\pm1.6$ & $17.6\pm5.5$ & $7.4\pm1.6$ & $13.7\pm7.8$ & $45.2\pm2.1$  \\ \hline
160 & $111$ & $101.7\pm6.8$  & $82.6\pm5.4$   & $10.5\pm1.4$ & $3.0\pm1.5$  & $2.2\pm0.4$ & $3.4\pm3.4$  & $122.9\pm5.6$ \\ \hline
200 & $159$ & $140.8\pm6.8$  & $108.2\pm4.5$  & $23.3\pm3.1$ & $3.4\pm1.5$  & $3.2\pm0.3$ & $2.7\pm3.7$  & $48.8\pm2.2$  \\ \hline
400 & $109$ & $110.8\pm5.8$  & $59.8\pm2.7$   & $35.9\pm4.7$ & $5.5\pm1.8$  & $9.3\pm1.1$ & $0.2\pm0.2$  & $17.5\pm0.8$  \\ \hline
\multicolumn{9}{|c|}{1-jet category} \\
\hline
120 & $72$  & $59.5\pm5.9$   & $27.0\pm4.7$   & $17.2\pm1.0$ & $5.4\pm2.4$  & $3.2\pm0.6$ & $6.6\pm2.3$  & $6.5\pm0.3$  \\ \hline
130 & $105$ & $79.9\pm7.7$   & $38.5\pm6.6$   & $25.6\pm1.4$ & $6.5\pm2.5$  & $4.0\pm0.6$ & $5.3\pm2.5$  & $17.6\pm0.8$ \\ \hline
160 & $86$  & $70.8\pm6.0$   & $33.7\pm5.5$   & $27.9\pm1.4$ & $3.2\pm1.4$  & $1.9\pm0.3$ & $4.2\pm1.4$  & $60.2\pm2.6$ \\ \hline
200 & $111$ & $130.8\pm6.7$  & $49.3\pm2.2$   & $59.4\pm2.8$ & $5.2\pm1.8$  & $2.2\pm0.1$ & $14.6\pm5.3$ & $25.8\pm1.1$ \\ \hline
400 & $128$ & $123.6\pm5.3$  & $44.6\pm2.2$   & $60.6\pm2.9$ & $6.2\pm2.1$  & $3.9\pm0.5$ & $8.3\pm3.2$  & $12.2\pm0.5$ \\ \hline
\multicolumn{9}{|c|}{2-jet category} \\
  \hline
120 & $8$   & $11.3\pm3.6$   & $1.3\pm0.2$    & $5.5\pm2.8$  & $0.7\pm0.6$  & $1.8\pm1.5$ & $1.9\pm1.4$ & $1.1\pm0.1$  \\ \hline
130 & $10$  & $13.3\pm4.0$   & $1.6\pm0.2$    & $6.5\pm3.2$  & $0.7\pm0.6$  & $1.8\pm1.5$ & $2.7\pm1.9$ & $2.7\pm0.2$  \\ \hline
160 & $12$  & $15.9\pm4.6$   & $1.9\pm0.2$    & $8.4\pm3.9$  & $1.2\pm0.8$  & $1.8\pm1.5$ & $2.7\pm1.9$ & $12.2\pm0.7$ \\ \hline
200 & $13$  & $17.8\pm5.0$   & $2.2\pm0.2$    & $9.4\pm4.2$  & $1.2\pm0.8$  & $1.8\pm1.5$ & $3.2\pm2.1$ & $8.4\pm0.5$  \\ \hline
400 & $20$  & $23.8\pm6.4$   & $3.5\pm0.3$    & $14.1\pm5.8$ & $1.1\pm0.8$  & $1.9\pm1.5$ & $3.3\pm2.1$ & $2.5\pm0.1$  \\ \hline
  \end{tabular}
  }
  \end{center}
\end{table*}

\section{Efficiencies and systematic uncertainties}
\label{sec:systematics}

The signal efficiency is estimated using simulations.
All Higgs production mechanisms are considered:
the gluon fusion process, the associated production of the Higgs boson with
a $\W$ or $\Z$ boson, and the VBF process. Since the Higgs $\pt$ spectrum generated by
\textsc{powheg} is harder than that predicted by more
precise calculations~\cite{Bozzi:2005wk,deFlorian:2011xf}, the Higgs boson $\pt$ distribution
is re-weighted to match the prediction from NNLO calculations with a resummation up to
next-to-next-to-leading-log accuracy, following the method proposed in~\cite{reweighting}.
The SM Higgs boson production cross sections are taken
from~\cite{LHCHiggsCrossSectionWorkingGroup:2011ti,Dawson:1990zj,Spira:1995rr,Harlander:2002wh,Anastasiou:2002yz,Tackmann:2011,Ravindran:2003um,Catani:2003zt,Actis:2008ug,Anastasiou:2008tj,deFlorian:2009hc,Ciccolini:2007jr,Ciccolini:2007ec,Arnold:2008rz,Brein:2003wg,Ciccolini:2003jy,hdecay2,Denner:2011mq,Bredenstein:2006rh,Bredenstein:2006ha}.

Residual discrepancies in the lepton reconstruction and identification
efficiencies between data and simulation are corrected for by
data-to-simulation scale factors measured using $\dyll$ events in the
$\Z$ peak region~\cite{wzxs}, recorded with dedicated unbiased triggers.
These factors depend on the lepton $\pt$ and $|\eta|$, and
are typically in the range (0.9-1.0).

Experimental effects, theoretical predictions, and the choice of Monte Carlo event
generators are considered as sources of uncertainty for both the cut-based and the BDT analyses.
For the cut-based analysis the impact of these uncertainties on the signal efficiency is assessed, while
for the BDT analysis the impacts on both the signal efficiency and the
kinematic distributions are considered.
The experimental uncertainties on lepton efficiency, momentum scale and resolution, $\met$
modeling, and jet energy scale are applied to the reconstructed objects in simulated events by smearing
and scaling the relevant observables and propagating the effects to the kinematic variables used
in the analysis. Separate $\Pq\Paq\to \WW$ samples are produced with varied
renormalization and factorization scales using the \textsc{mc@nlo} generator~\cite{MCatNLO} to address the
shape uncertainty in the theoretical model. The kinematic differences with respect to an
alternate event generator are used as an additional uncertainty for
$\Pq\Paq\to \WW$ (\textsc{madgraph} versus \textsc{mc@nlo}) and
top-quark production (\textsc{madgraph} versus \textsc{powheg}).
The normalization and the shape uncertainty on the $\Wjets$ background is included by varying
the efficiency for misidentified leptons to pass the tight lepton
selection and by comparing to the results of a closure test using simulated samples.
For the BDT analysis, the $\dyll$ process is modeled using events at low
$\met$ to gain statistical power in the extrapolation to the signal region. The effect of the
limited amount of simulated events on the shape knowledge is addressed by varying the
distribution used to set the limits by the statistical uncertainty in each histogram bin.

The uncertainty on the signal efficiency from pile-up is evaluated to be $0.5\%$.
The assigned uncertainty corresponds to shifting the mean of the
expected distribution which is used to reweight the simulation
up and down by one interaction.
A $4.5\%$ uncertainty is assigned to the luminosity
measurement~\cite{lumiPAS}.

The systematic uncertainties due to theoretical ambiguities are separated
into two components, which are assumed to be
independent. The first component is the uncertainty on the fraction of
events categorized into the different jet categories and the effect of jet bin migration.
The second component is the uncertainty on
the lepton acceptance and the selection efficiency of all other
requirements. The effect of variations in parton distribution functions and the
value of $\alpha_{s}$, and the effect of higher-order corrections,
are considered for both components using the \textsc{pdf4lhc}
prescription~\cite{Botje:2011sn,Alekhin:2011sk,Lai:2010vv,Martin:2009iq,Ball:2011mu}.
For the jet categorization, the effects of higher-order log terms via
the uncertainty in the parton shower model and the underlying event
are also considered, by comparing different generators. These uncertainties range
between 10\% and 30\% depending on the jet category.
The uncertainties related to the diboson
cross sections are calculated using the \textsc{mcfm} program~\cite{MCFM}.

The overall signal efficiency uncertainty is estimated to be about 20\%
and is dominated by the theoretical uncertainty due to missing
higher-order corrections and PDF uncertainties. The uncertainty on the background estimations
in the $\hww$ signal region is about 15\%, which is dominated by the
statistical uncertainty on the observed number of events in the background-control regions.

\section{Results}
\label{sec:results}
After applying the mass-dependent Higgs selection, no significant
excess of events is found with respect to the expected backgrounds,
and upper limits are derived on the product of the Higgs boson
production cross section and the $\Hi \to \WW$ branching fraction,
$\sigma_{\Hi} \times \mathrm{BR}(\Hi \to \WW)$, with respect to the
SM Higgs expectation, $\sigma/\sigma_\text{SM}$.

\begin{figure}[htbp]
  \begin{center}
   \includegraphics[width=\cmsFigWidth]{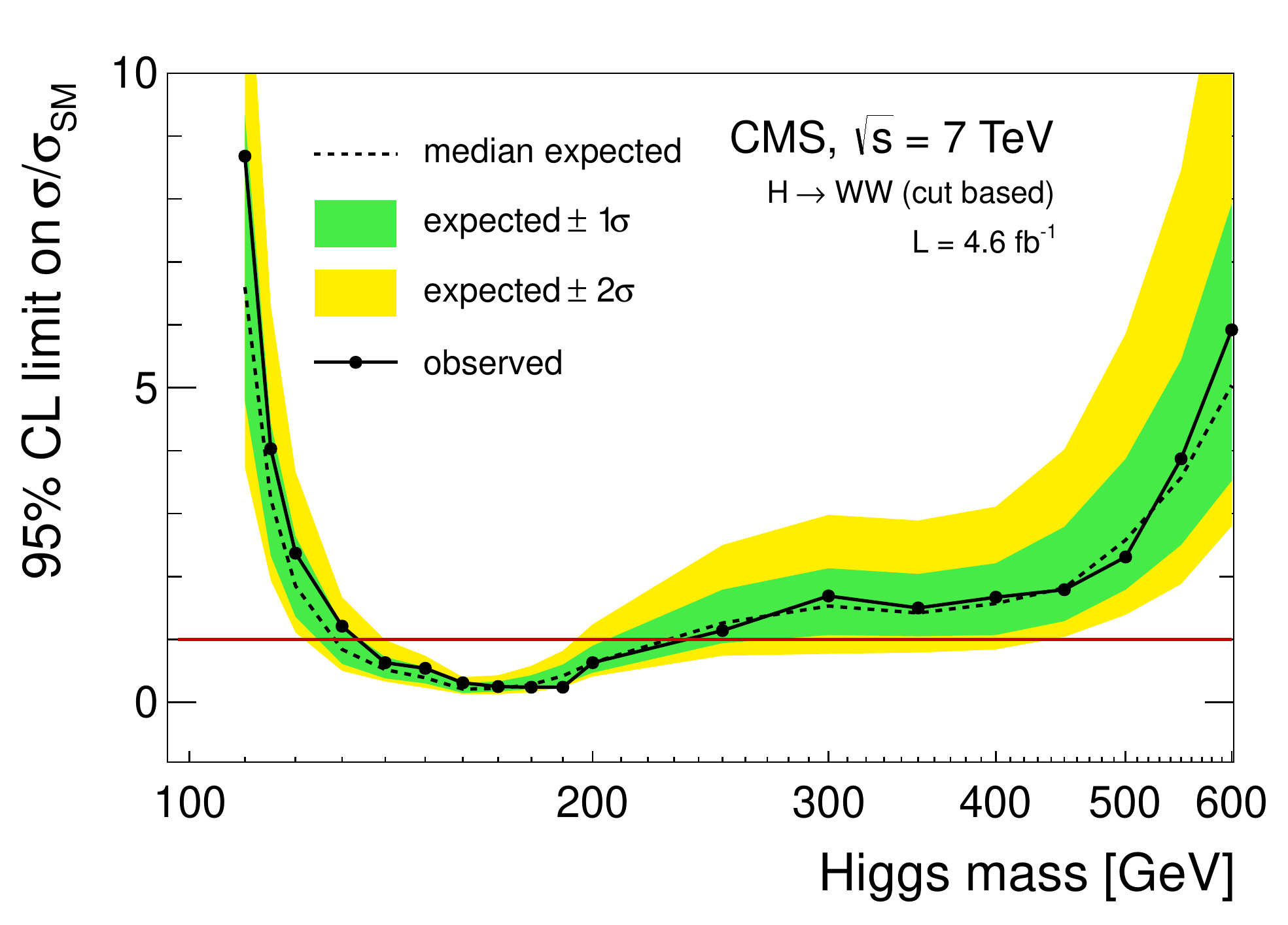}
   \includegraphics[width=\cmsFigWidth]{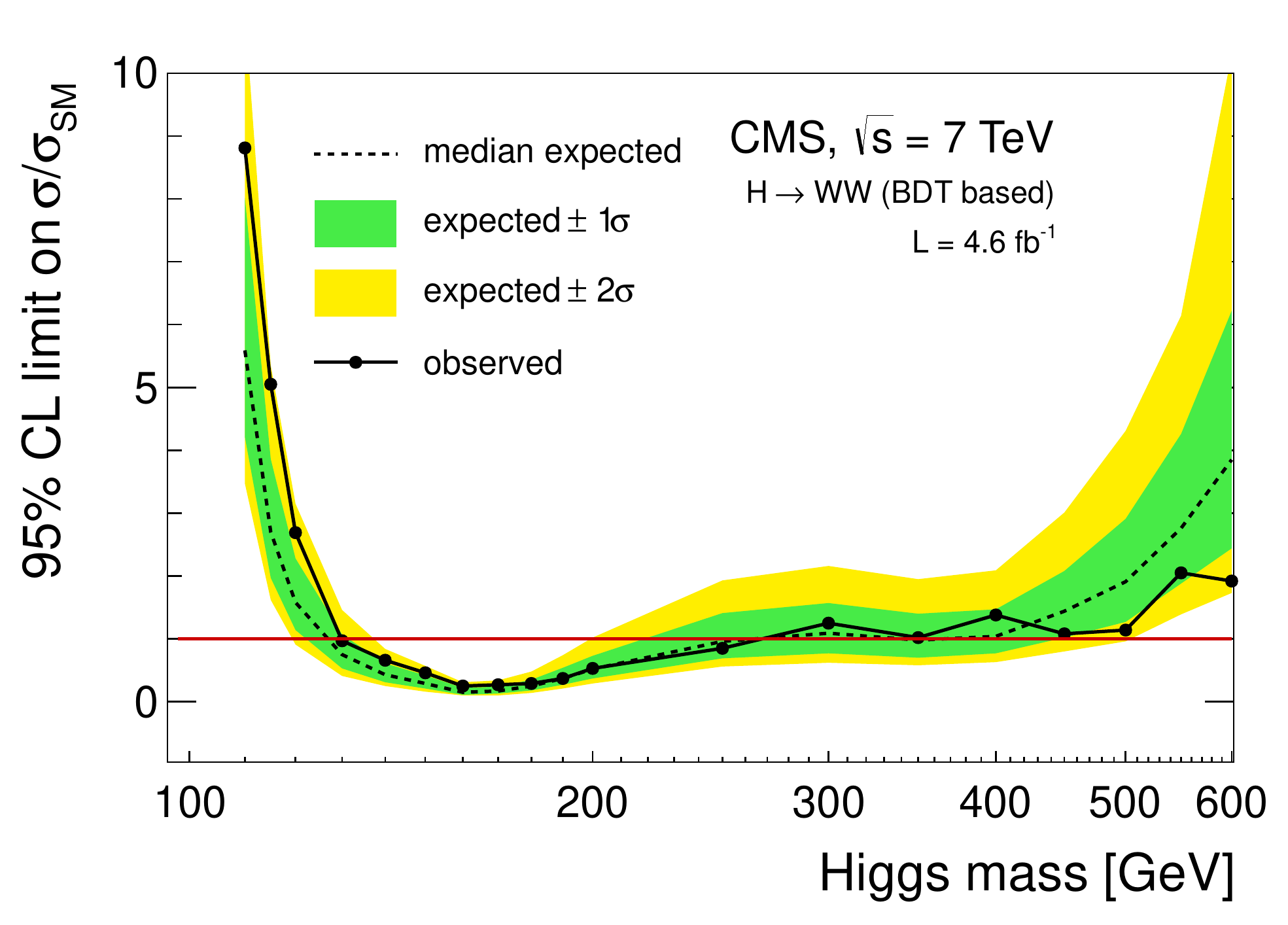}
    \caption{Expected and observed 95\% CL upper
       limits on the cross section times branching fraction,
       $\sigma_{\PH} \times \mathrm{BR}(\PH \to \WW)$,
       relative to the SM Higgs expectation, using cut-based (\cmsLeft) and
       multivariate BDT (\cmsRight) event selections. Results are obtained using the CL$_{s}$ approach.}
    \label{fig:xsLimCuts}
  \end{center}
\end{figure}

To compute the upper limits the modified frequentist construction
CL$_{s}$~\cite{Read1,junkcls,LHC-HCG} is used. The likelihood function from the
expected number of observed events is modeled as a Poisson random variable,
whose mean value is the sum of the contributions from signal and background
processes. All the sources of systematic uncertainties are also considered.
The 95\% CL observed and expected median upper limits are shown in
Fig.~\ref{fig:xsLimCuts}. Results are reported for both the
cut-based and the BDT approaches. The bands represent the $1 \sigma$ and
$2 \sigma$ probability intervals around the expected limit. The
\textit{a posteriori} probability intervals on the cross section are
constrained by the assumption that the signal and
background cross sections are positive definite.

The cut-based analysis excludes the presence of a Higgs boson
with mass in the range 132--238\GeV at 95\% CL, while the
expected exclusion limit in the hypothesis of background only
is 129--236\GeV. With the multivariate analysis,
a Higgs boson with mass in the range
129--270\GeV is excluded at 95\% CL, while the expected exclusion limit for the
background only hypothesis is in the range 127--270\GeV.
The observed (expected) upper limits are about 0.9 (0.7) times
the SM expectation for $\mHi=130\GeV$.

\section{Summary}
\label{sec:summary}
A search for the SM Higgs boson decaying to $\WW$ in pp
collisions at $\sqrt{s} = 7\TeV$ is performed by the CMS experiment
using a data sample corresponding
to an integrated luminosity of $\usedLumi$. No significant excess of events
above the SM background expectation is found. Limits on
the Higgs boson production cross section relative to the SM Higgs
expectation are derived, excluding the presence of the SM Higgs boson
with a mass in the range 129--270\GeV at 95\% CL.

\section*{Acknowledgements}
We wish to congratulate our colleagues in the CERN accelerator departments for the excellent
performance of the LHC machine. We thank the technical and administrative staff at CERN and other CMS
institutes, and acknowledge support from: FMSR (Austria); FNRS and FWO (Belgium); CNPq, CAPES, FAPERJ,
and FAPESP (Brazil); MES (Bulgaria); CERN; CAS, MoST, and NSFC (China); COLCIENCIAS (Colombia); MSES
(Croatia); RPF (Cyprus); Academy of Sciences and NICPB (Estonia); Academy of Finland, MEC, and HIP
(Finland); CEA and CNRS/ IN2P3 (France); BMBF, DFG, and HGF (Germany); GSRT (Greece); OTKA and NKTH
(Hungary); DAE and DST (India); IPM (Iran); SFI (Ireland); INFN (Italy); NRF and WCU (Korea); LAS
(Lithuania); CINVESTAV, CONACYT, SEP, and UASLP-FAI (Mexico); MSI (New Zealand); PAEC (Pakistan);
MSHE and NSC (Poland); FCT (Portugal); JINR (Armenia, Belarus, Georgia, Ukraine, Uzbekistan); MON,
RosAtom, RAS and RFBR (Russia); MSTD (Serbia); MICINN and CPAN (Spain); Swiss Funding Agencies
(Switzerland); NSC (Taipei); TUBITAK and TAEK (Turkey); STFC (United Kingdom); DOE and NSF (USA).
Individuals have received support from the Marie-Curie programme and the European Research Council
(European Union); the Leventis Foundation; the A. P. Sloan Foundation; the Alexander von Humboldt
Foundation; the Belgian Federal Science Policy Office; the Fonds pour la Formation \`a la Recherche
dans l'\'industrie et dans l'\'Agriculture (FRIA-Belgium); the Agentschap voor Innovatie door Wetenschap
en Technologie (IWT-Belgium); the Council of Science and Industrial Research, India; and the HOMING
PLUS programme of Foundation for Polish Science, cofinanced from European Union, Regional Development
Fund.

\bibliography{auto_generated}   
\cleardoublepage \appendix\section{The CMS Collaboration \label{app:collab}}\begin{sloppypar}\hyphenpenalty=5000\widowpenalty=500\clubpenalty=5000\input{HIG-11-024-authorlist.tex}\end{sloppypar}
\end{document}

%% file: HIG-11-024-authorlist.tex
\textbf{Yerevan Physics Institute,  Yerevan,  Armenia}\\*[0pt]
S.~Chatrchyan, V.~Khachatryan, A.M.~Sirunyan, A.~Tumasyan
\vskip\cmsinstskip
\textbf{Institut f\"{u}r Hochenergiephysik der OeAW,  Wien,  Austria}\\*[0pt]
W.~Adam, T.~Bergauer, M.~Dragicevic, J.~Er\"{o}, C.~Fabjan, M.~Friedl, R.~Fr\"{u}hwirth, V.M.~Ghete, J.~Hammer\cmsAuthorMark{1}, M.~Hoch, N.~H\"{o}rmann, J.~Hrubec, M.~Jeitler, W.~Kiesenhofer, M.~Krammer, D.~Liko, I.~Mikulec, M.~Pernicka$^{\textrm{\dag}}$, B.~Rahbaran, C.~Rohringer, H.~Rohringer, R.~Sch\"{o}fbeck, J.~Strauss, A.~Taurok, F.~Teischinger, P.~Wagner, W.~Waltenberger, G.~Walzel, E.~Widl, C.-E.~Wulz
% ...
\vskip\cmsinstskip
\dag:~Deceased\\
1:~~Also at CERN, European Organization for Nuclear Research, Geneva, Switzerland\\
% ...
